\documentclass[aps,pra,twocolumn,nopacs,superscriptaddress,groupedaddress]{revtex4}

\usepackage{amssymb}
\usepackage{color,graphicx}
\usepackage{amsmath}
\usepackage{amsbsy}
\usepackage{amsthm}
\usepackage{bbm}
\usepackage{bm}
\usepackage{epsfig}
\usepackage{lscape}
\usepackage{float}
\usepackage{subfigure}

\newcommand{\D}{\text{d}}

\begin{document}

\title{Testing Dissipative Collapse Models with a Levitated Micromagnet}

\author{A. Vinante}
\email{andrea.vinante@ifn.cnr.it}
\affiliation{School of Physics and Astronomy, University of Southampton, SO17 1BJ, Southampton, UK}
\affiliation{Istituto di Fotonica e Nanotecnologie - CNR and Fondazione Bruno Kessler, I-38123 Povo, Trento, Italy}
\author{G. Gasbarri}
\altaffiliation[Now at: ]{Universitat Aut\`{o}noma de Barcelona}
\affiliation{School of Physics and Astronomy, University of Southampton, SO17 1BJ, Southampton, UK}
\author{C. Timberlake}
\affiliation{School of Physics and Astronomy, University of Southampton, SO17 1BJ, Southampton, UK}
\author{M. Toro\v{s}}
\affiliation{Department of Physics and Astronomy, University College London, Gower Street, London WC1E 6BT, UK}
\author{H. Ulbricht}
\email{h.ulbricht@soton.ac.uk}
\affiliation{School of Physics and Astronomy, University of Southampton, SO17 1BJ, Southampton, UK}

\date{\today}

\begin{abstract}
We present experimental tests of dissipative extensions of spontaneous wave function collapse models based on a levitated micromagnet with ultralow dissipation. The spherical micromagnet, with radius $R=27$ $\mu$m, is levitated by Meissner effect in a lead trap at $4.2$ K and its motion is detected by a SQUID. We perform accurate ringdown measurements on the vertical translational mode with frequency $57$ Hz, and infer the residual damping at vanishing pressure $\gamma/2\pi<9$ $\mu$Hz. From this upper limit we derive improved bounds on the dissipative versions of the CSL (continuous spontaneous localization) and the DP (Di\'{o}si-Penrose) models with proper choices of the reference mass. In particular, dissipative models give rise to an intrinsic damping of an isolated system with the effect parameterized by a temperature constant; the dissipative CSL model with temperatures below 1 nK is ruled out, while the dissipative DP model is excluded for temperatures below $10^{-13}$ K. Furthermore, we present the first bounds on dissipative effects in a more recent model, which relates the wave function collapse to fluctuations of a generalized complex-valued spacetime metric.    
\end{abstract}

\maketitle

Spontaneous wave function collapse models~\cite{GRW,CSL,DP1,DP2,collapse_review1,collapse_review2} are a well established approach in the context of quantum foundations. The key idea is that the unitary evolution of standard quantum mechanics must be modified by additional phenomenological terms in order to explain the emergence of definite and stochastic outcomes in measurement processes. These additional terms must be nonlinear and stochastic, leading to a fundamental breaking of the quantum superposition principle. In a nutshell, collapse models postulate the existence of some kind of classical noise field, the nature of which is either unknown \cite{GRW,CSL} or related to a cosmological or to the gravitational field \cite{DP1,DP2}. It has also been suggested that collapse models could be related to the long standing problem of the incompatibility between quantum mechanics and general relativity \cite{stamp}. In this latter respect, other related phenomenological models have been recently investigated \cite{adlerCG,bassiCG,sudarsky,tilloy1}. 

Collapse models are usually parameterized by only a few free parameters: a collapse rate which sets the strength of the collapse mechanism, and a localization length which quantifies the localization precision. The parameters of the model can be considered as independent, as in the continuous spontaneous localization (CSL) model \cite{CSL}, or fixed by theoretical considerations, e.g. the collapse rate in the Di\'{o}si-Penrose (DP) model which is set by gravity. 

A well-known issue of collapse models is the energy divergence problem: the collapse noise feeds continuously energy into any material system, implying an unbounded rate of increase of energy in the universe \cite{GRW}. This problem is solved by dissipative extensions of collapse models, in which the noise is associated to a dissipative mechanism and can thus be thought as a thermal bath interacting with ordinary matter \cite{smirne,smirne1}. In this framework, the energy can flow in both directions and will not diverge with time anymore. Dissipative models imply the existence of a fundamental and universal damping mechanism which can be in principle probed by mechanical systems with very low dissipation \cite{pontin}.

In this paper we perform new experimental tests of the dissipative versions of the continuous spontaneous localization (CSL) model \cite{smirne, nobakht} and the DP model \cite{smirne1,pontin}, also known as dCSL and dDP. Our experiment is based on a magnetically levitated microsphere with ultralow damping. In particular, our data exclude a new portion of the parameter space compared to previous experiments \cite{pontin} substantially excluding collapse temperatures lower than $10^{-9}$ K for the dCSL model and $10^{-13}$ K for the dDP model, with proper choices of the reference mass. In addition, we test for the first time a more recent model first proposed by Adler \cite{adlerCG,gasbarritoros}, which assumes the collapse noise to arise from complex fluctuations of the gravitational field, or equivalently of the spacetime metric. We refer to this model as CGF (complex gravity fluctuations). We show that our data allow to probe complex fluctuations of the metric with an imaginary part down to $10^{-22}$. 

\section{Theory}

\subsection{The dCSL model}

CSL, the most studied among collapse models, is constructed in such way to produce a spatial localization of the wavefunction, i.e. a collapse in position. The localization rate scales with the mass of the system, implying a rapid collapse of the center-of-mass position of any macroscopic system, while giving no measurable effect at the microscopic level, where conventional quantum mechanics is recovered. The standard CSL model has two free parameters, the collapse rate conventionally referred to a single nucleon $\lambda$, and a characteristic length $r_c$. Many different experimental techniques have been recently proposed or implemented to test the CSL model. While true interferometric tests have recently achieved impressive sensitivity \cite{toros2017, arndt}, even more stringent bounds on the CSL collapse rate $\lambda$ have been established by non-interferometric experiments looking at noise and diffusion in mechanical systems \cite{vinanteCSL1,vinanteCSL2, vinanteCSL3, lisa,helou,cinesi} or cold atoms \cite{bilardello}, or in spontaneous generation of x-ray photons \cite{xray} or high frequency phonons \cite{adlervinante}.

The dCSL model has been explicitly introduced to remove the energy divergence of the standard CSL model \cite{smirne}. Formally, in dCSL the collapse happens both in position and in momentum \cite{smirne}. The evolution of the density matrix of the center-of-mass of a rigid body along a fixed direction $x$ is described, in the limit of small $x$ and $p$, by the following Lindblad-type master equation \cite{nobakht}:
\begin{align} \label{master}
\begin{aligned}
\frac{\D \hat{\rho}}{\D t} =& -\frac{i}{\hbar}[\hat H, \hat \rho ] -\frac{\eta}{2}[\hat x,[\hat x, \hat \rho]] -\frac{\gamma_c^2}{8 \eta \hbar^2}[\hat p,[\hat p, \hat \rho]] \\
&-\frac{i \gamma_c}{2 \hbar}  [\hat x, \lbrace \hat p, \hat \rho \rbrace ],
\end{aligned}
\end{align}
where $\hat H$ is the standard Hamiltonian, the second and third term on the right hand side describes position and momentum decoherence/diffusion due to the dCSL effect and the fourth one accounts for dCSL energy dissipation. 

Under the assumption $r_c \gg a$ \cite{note} with $a$ the interatomic distance, the diffusion parameter $\eta$ can be expressed as a function of the free parameters of the model and the mass distribution of the rigid body \cite{nobakht,pontin}:
\begin{align}  \label{eta}
 \eta &=\frac{(4\pi)^{\frac{3}{2}}\,\lambda\,r_c^3}{\hbar^2 m_0^2 (2 \pi \hbar)^3}\,
 \int\D{\bf q}\,|  \tilde{\varrho}({\bf q})|^2\,e^{-\frac{{\bf q}^2 r_c^2 \left( 1+ \chi \right)^2}{\hbar^3}}\,q^2_x
\end{align}
with $m_0$ the nucleon mass, ${\bf q}=(q_x,q_y,q_z)$ the momentum, $\varrho({\bf r})$ the mass density in the coordinate space and
\begin{equation}
\tilde{\varrho}({\bf q})=\int\D{\bf r}\,e^\frac{i{\bf q}\cdot{\bf r}}{\hbar}\,\varrho({\bf r})
\end{equation}
its Fourier transform.

The free parameters of the model are the collapse rate $\lambda$, the characteristic length $r_c$, and the dimensionless dissipation parameter $\chi$. The latter can be rewritten in terms of a new parameter $T_c$ in the following way \cite{pontin}:
\begin{equation} \label{chi}
\chi=\frac{\hbar^2}{8 m_a r_c^2 k_B T_{c}}.   \end{equation}
$T_c$ can be interpreted as the temperature 
of the collapse field \cite{smirne}.
The energy dissipation rate of the center-of-mass dynamics can be written as \cite{smirne,nobakht}:
\begin{equation}
 \gamma_c =4 \eta r_c^2 \chi \left( 1+ \chi \right)\frac{m_a}{m}   \label{gamma}
\end{equation}
where $m$ is the total mass and $m_{a}$ is a reference mass for the elementary entity which constitutes the physical object. Following Refs.~\cite{pontin,bilardello}, we choose $m_a$ to be the nuclear mass. We underline that the reference mass $m_a$ for the dCSL model does not necessarily coincide with the nucleon mass $m_0$, which is the usual convention for the standard CSL model. For the latter case, choosing a different value for $m_0$ only amounts to a redefinition of the collapse rate without affecting the predictions of the model for, say, a crystal of a mass $m$. More physically relevant is however the nuclear mass, which can be taken as the physical unit reference for the crystal structure. The non-dissipative CSL model takes implicitly into account the nuclear mass through the amplification mechanism which rescales the reference mass $m_0$ to the nuclear mass for $r_c>1$ fm. For the dissipative model the second reference mass is more physically relevant because it is related to the temperature $T_c$ through Eq.~(\ref{chi}). By setting $m_a$ equal to the nuclear mass, as done in Refs.~\cite{pontin,bilardello}, we implicitly assume that the internal dynamics of the nuclei is irrelevant for the dissipative features of the dCSL model, similarly as for the non-dissipative features. This assumption is ultimately justified by the the fact that $r_c$ is much larger than the nuclear size. 

As one may notice, the standard CSL is recovered when $\chi=0$, that according to Eq.~(\ref{chi}) corresponds to a CSL field with infinite temperature. Technically, this implies an energy divergence, as the CSL noise will continuously transfer energy to the system causing an unbounded momentum diffusion. This unpleasant consequence is removed in the dissipative version. Indeed, an isolated system will eventually thermalize to $T_c$ \cite{smirne}, meaning that for temperatures higher than $T_c$ the dCSL noise will effectively act as a refrigerator. The proponents of the dCSL model further propose that reasonable values for $T_c$ should be around $1$ K, by analogy to other known cosmological fields such as cosmic microwave photons or cosmic neutrinos \cite{smirne}. Concerning the other parameters two main proposals are known in literature for the CSL model, the initial guess by Ghirardi {\it et al.} who proposed $\lambda \approx 10^{-16}$ Hz at $r_c=10^{-7}$ m \cite{GRW} and the one by Adler, who proposed a much higher value $\lambda \approx 10^{-8\pm 2}$ Hz at $r_c=10^{-7 }$ m, motivated by making the collapse effective at mesoscopic scale \cite{adler}.

For the case relevant to our work, namely a homogeneous sphere of radius $R$ and density $\varrho$, the Fourier transform of the mass density is
\begin{align}
\tilde {\varrho}(q)= \frac{3 m \hbar}{q R} J_{1}(q R /\hbar),   \label{rhotilde}
\end{align}
where $J_{i}$ represents the $i$-th spherical Bessel function. The integral in Eq.~(\ref{eta}) can be analytically solved \cite{pontin} providing the diffusion constant and dissipation rate:
\begin{align}
 \eta &= \frac{3 \lambda m^2 r_c^2}{\left( 1+\chi \right) m_0^2 R^4 } K\left[ \frac{R}{r_c \left(1+\chi \right)} \right] , \label{eta1} \\
 \gamma_c &= \frac{3 \lambda \hbar^2 m m_a r_c^2}{2 k_B T_c m_0^2 R^4} K\left[ \frac{R}{r_c \left(1+\chi \right)} \right]  ,    \label{gamma1}
\end{align}
where we have defined for convenience:
\begin{equation}
K \left( y \right) = 1-\frac{2 }{y^2} + e^{-y^2} \left( 1+ \frac{2}{y^2} \right) .
\end{equation}
The function $K(y)$ can be approximated by 1 and $y^4/6$ for large and small $y$, respectively. This determines the behaviour of $\eta$ and $\gamma_c$ as function of $r_c$. In the limit of small dissipation $\chi \ll 1$ both functions are proportional to $r_c^2$ if $r_c \ll R$ and to $r_c^{-2}$ if $r_c \gg R$, with a shallow maximum at $r_c \approx R$. This picture breaks down for very low $T_c$, such that $\chi>1$. In this limit both diffusion and dissipation feature a stronger dependence on $r_c$ for small $r_c$, corresponding to $\eta \propto r_c^{8}$ and $\gamma_c \propto r_c^{6}$, respectively.

\subsection{The dDP model}
The fact that the collapse effect scales with the mass of the system suggests a natural connection to gravity. The Di\'{o}si-Penrose (DP) model \cite{DP1,DP2} is an attempt to provide this link. Although proposed by  Di\'{o}si \cite{DP1}, the model is known in literature as DP because it captures some features of a related proposal by Penrose \cite{DP2}. The master equation of the DP model is almost identical to the one of CSL, only differing from the latter in the localization operator. However in the DP model the collapse strength is set proportional to the gravitational constant G rather than depending on a free parameter. As such, the standard DP model features only one free parameter, a regularization length $R_0$ \cite{bassiCG}. Proposed values for $R_0$ range from $10^{-15}$~m \cite{DP1} to $10^{-7}$~m \cite{GGR}. 

A dissipative extension of the DP model (so called dDP model) can be developed in a similar way as the dCSL \cite{smirne1,pontin}. One defines a dissipation parameter $\chi_{\text{DP}}$, which can be rewritten in terms of a 
collapse field temperature $T_{\text{DP}}$:
\begin{equation} \label{chiDP}
\chi_{\text{DP}}=\frac{\hbar^2}{8 m_a R_0^2 k_B T_{\text{DP}}}. 
\end{equation}
In the limit of uniform mass density in which the characteristic length $R_0'=R_0 \left( 1+ \chi_{DP} \right)$ is larger than the interatomic distance $a$, the expression for the diffusion constant for a homogeneous sphere was calculated in \cite{pontin} as:
\begin{equation}
\eta_{\text{DP}} = \frac{G m^2 }{\sqrt{\pi} R^3} I \left( \frac{R}{R_0'} \right)   \label{etaDP1}
\end{equation}
where we have defined for convenience:
\begin{equation}
I(y) = \sqrt{\pi} \mathrm{Erf}(y)+ \frac{1}{y} \left( e^{-y^2}-3 \right)+\frac{2}{y^3} \left(  1- e^{-y^2} \right).
\end{equation}
The dissipation rate $\gamma_{\text{DP}}$ is then calculated as:
\begin{equation}
 \gamma_\text{DP} =4 \eta R_0^2 \chi_{\text{DP}} \left( 1+ \chi_{\text{DP}} \right)\frac{m_a}{m}.   \label{gammaDP}
\end{equation} 
The function $I(y)$ can be approximated by $y^3/6$ for $y \ll 1$ and tends to $\sqrt{\pi}$ for $y \gg 1$. Therefore, the collapse/diffusion parameter $\eta_\text{DP}$ scales with $m^2$ for $R < R_0'$ and with $m$ for $R>R_0'$. This behaviour is typical of collapse models, and can be interpreted as a coherent amplification of the collapse rate within a sphere of radius $R_0'$ \cite{toros}.  

In the opposite limit, i.e. when $R’_{0} \ll a$ the assumption of homogeneity is no longer valid and we need to consider the granularity of the matter distribution.
In this regime the diffusion parameter $\eta_{DP}$ can be calculated by means of a lattice model and the following expression obtained:
\begin{equation}
\eta_{\text{DP}}=\frac{G}{6\hbar\sqrt{\pi}(1+\chi_\text{DP})^{3}R_{0}^{3}}m_{a}m , \label{etaDP2}
\end{equation}
The attentive reader can find  detailed calculations in Appendix. We note that, in contrast with the CSL model, the granular limit for the DP model is often suggested because it allows to enhance the collapse rate, making it closer to experimental testability \cite{bassiCG}. $R_0$ as low as the nuclear size has been proposed in literature -- indeed the DP model is nonrelativistic and the nucleon scale is a natural limit for the nonrelativistic regime \cite{bassiCG}.



We note that in the dDP model there are only two free parameters, the regularization length $R_0$ and the collapse field temperature $T_{\text{DP}}$. Similarly as in dCSL, a system will eventually thermalize to the temperature $T_{\text{DP}}$, while for the standard DP model there is no dissipation, leading to an energy divergence.

\subsection{The CGF model}
The complex gravity fluctuations (CGF) model is based on assuming the existence of complex fluctuations of the gravitational field, or equivalently of the spacetime metric. The idea, first proposed by Adler \cite{adlerCG} and further developed in Refs. \cite{gasbarritoros} and \cite{bassiCG}, can be summarized as follows. 

A gravitational field $h_{\mu\nu}$ couples to the  stress energy tensor $T_{\mu\nu}$ of the system. In a linearized fully quantum theory \cite{blencowe} this implies the existence of a coupling term $H_{\text{int}}= \frac{1}{2}h_{\mu\nu}T^{\mu\nu}$ in the Hamiltonian, that in the non relativistic regime can be simplified as $H_{\text{int}}=\frac{1}{2} h_{00}mc^{2}$. This ultimately leads to a master equation of decoherence type such as Eq.~(\ref{master}). However, if one assumes that the metric remains classical, but involves rapidly fluctuating complex terms, the resulting classical noise field would feature an antihermitian coupling to matter \cite{gasbarritoros,bassiCG}, which is the basic ingredient required to produce the collapse/localization of the wavefunction, as opposed to quantum decoherence. While in general relativity the metric is rigorously real-valued, complex effective metrics have been actually proposed in some modified gravity theories with chiral deformations \cite{krasnov}.


The noise-matter coupling in the case of classical complex noise will also lead to the appearance of nonlinear terms in the master equation. The derivation of the appropriate master equation for the center of mass of mechanical oscillator is reported in Ref.~\cite{gasbarritoros}. 
Here, we rewrite Eq.~(D5) in Ref.~\cite{gasbarritoros} as:
\begin{align} \label{eq:mastergc}
\begin{aligned}
\partial_{t}\hat{\rho}=&\frac{i}{\hbar}[\hat{H}_{0},\hat \rho]-\eta_{\text{CGF}} \left[ \hat{x},\left[ \hat x,\hat{\rho}\right]\right] +\frac{\gamma_{\text{CGF}}^{\text{R}}}{2 \hbar}  \left[ \hat{x},\left[ \hat{p},\hat{\rho}\right]\right] \\
&-i\frac{\gamma_{\text{CGF}}^{\text{I}}}{2 \hbar} \left[ \hat{x},\left\{ \hat{p},\hat{\rho} \right\}\right]
\end{aligned}
\end{align}
with $\hat{H}_{0}$ the Hamiltonian characterising the harmonic oscillator free dynamics, and
\begin{align}
\eta_{\text{CGF}} &=\frac{c^{4}\xi^{2}}{6 \pi^{2}\hbar^{7}}\int_{0}^{\infty} d q \int d\tau\, D^{\text{R}}(q,\tau)\tilde{\varrho}(q)^{2} q^{4} \\
\gamma_{\text{CGF}}^{\text{I/R}}&=\frac{c^{4}\xi^{2}}{3 \pi^{2}\hbar^{6} m}\int_{0}^{\infty} d q \int d\tau\,\tau D^{\text{I/R}}(q,\tau)\tilde{\varrho}(q)^{2} q^{4}   \label{eq:gamma}
\end{align}
where $D^{\text{I/R}}(q,\tau)$ are the real and imaginary part of the normalized correlator of the complex metrics fluctuations, expressed as function of time $\tau$ and momentum $q$, and $\xi$ is the dimensionless magnitude of the correlator. The dissipative term, with energy dissipation rate $\gamma_{\text{CGF}}^{\text{I}}$, depends only on the imaginary part of the correlator, while the real part leads to diffusion.
To proceed we assume that the imaginary part of correlator can be written as $D^{\text{I}}(q,\tau)=f(\tau)d(q)$ with 
\begin{align}
f(\tau)&= e^{-\lambda \tau}\nonumber\\
d(q)&=r_{c}^{3}\,e^{-r_{c}^{2}q^{2}/\hbar^{2}}
\end{align}
so to have $D(r,\tau)$ dimensionless and characterized by Gaussian spatial correlation with width $r_c$ as in the CSL model, and a time correlation with single exponential parameter $\lambda$. 
By inserting the mass density Eq.~(\ref{rhotilde}) and carrying out the integration we obtain
\begin{align}
&\int_{0}^{\infty} d\tau\, \tau e^{-\lambda \tau} = \lambda^{-2} \nonumber\\
&\int_{0}^{q} q^{4} d(q) |\tilde{\varrho}(q)|^{2} = \frac{9 r_{c}^{2}\hbar^{5} m^{2}\sqrt{\pi}}{4 R^{4}} K \left( \frac{R}{r_c} \right)
\end{align}
and combining the results together in Eq.~\eqref{eq:gamma} we find:
\begin{align}
\gamma_{\text{CGF}}^{\text{I}}= \frac{6 r_{c}^{2} c^{2}\xi^{2}}{(4\pi)^{\frac{3}{2}} R^4 \lambda^{2}}\frac{m c^{2}}{\hbar} K \left(  \frac{R}{r_c}\right).  \label{gammaCGF}
\end{align}
Note that, due to the assumption of gaussian spatial correlation, Eq.~(\ref{gammaCGF}) is very similar to the expression of $\gamma_c$ in the dCSL case Eq.~(\ref{gamma1}).
If we further assume that temporal and spatial correlations are related to each other by the speed of light and set $\lambda= c/r_{c}$, we obtain:
\begin{align}
\gamma_{\text{CGF}}^{\text{I}}= \frac{6 m c^{2}\xi^{2}}{(4\pi)^{\frac{3}{2}}\hbar}\frac{r_{c}^{4}}{R^{4}} K \left(  \frac{R}{r_c}\right).  \label{gammaCGF2}
\end{align}
This further assumption is suggested by the fact that the gravitational field propagates at speed of light \cite{gasbarritoros}.

\section{Experimental setup} 
Experimentally, we follow the approach outlined in Ref.~\cite{pontin}. By preparing and measuring a mechanical resonator with very low friction it is possible to set an upper bound on the fundamental dissipation predicted by collapse models. The advantage of this approach is that measuring very low dissipation is experimentally less challenging than measuring noise. In fact, ultralow mechanical dissipation is more easily achieved at low frequencies, where excess vibrational noise of seismic or acoustic origin are ubiquitous and very hard to shield.
\begin{figure}[!ht]
\includegraphics[width=8.6cm]{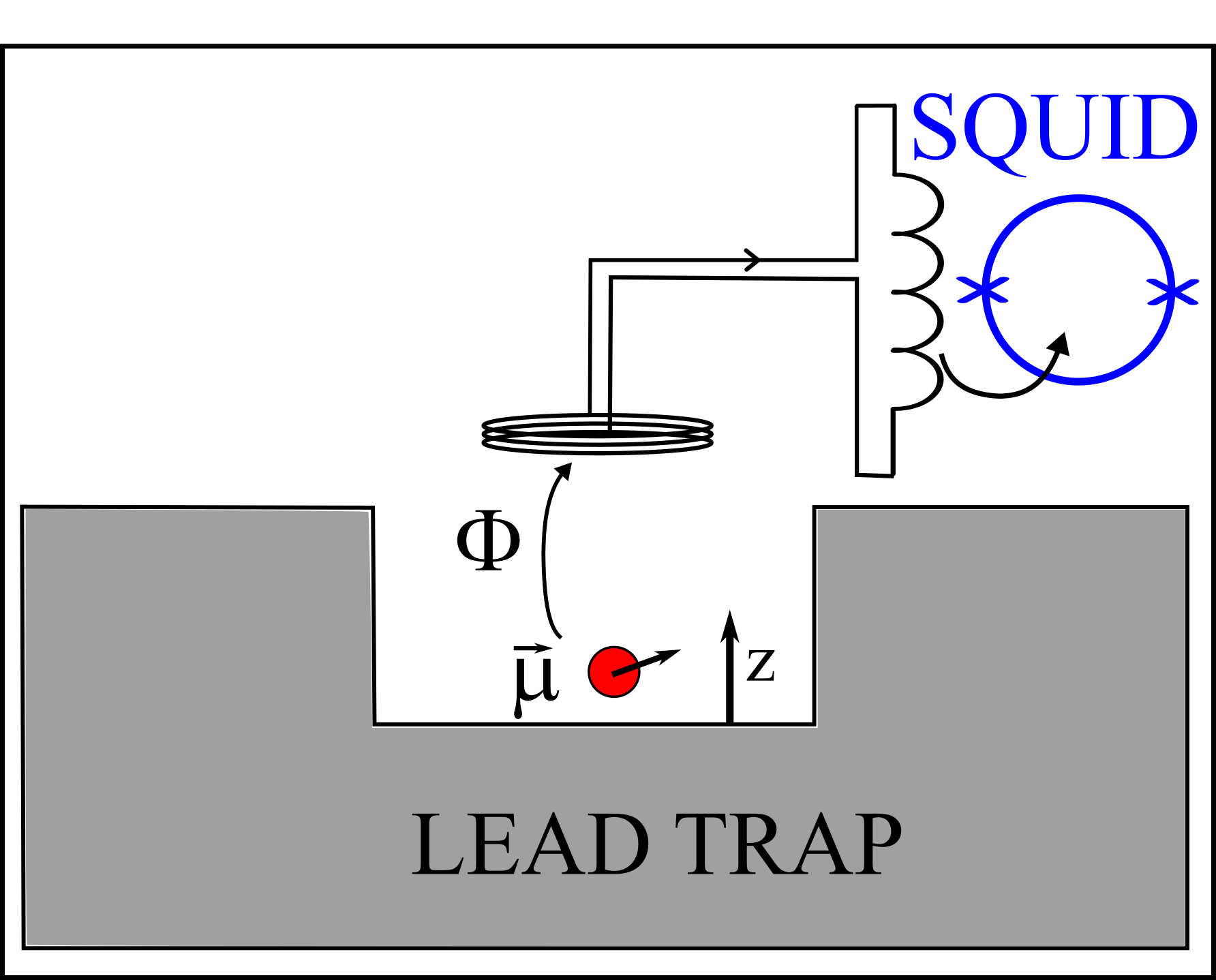}
\caption{Simplified scheme of the experimental setup. A micromagnetic sphere with permanent magnetic moment $\bm{\mu}$ is levitated by Meissner effect in a cylindrical well machined in a type-I superconductor (lead). The motion of the micromagnet is trapped in all degrees of freedom. For this work we consider specifically the vertical motion along $z$, which features a resonance frequency $f_0=56.8$ Hz. The motion is monitored by a commercial SQUID through the flux $\delta \Phi \propto \delta z$ induced in a superconducting pick-up coil.}  \label{scheme}
\end{figure}

Our mechanical resonator is a translational mode of a ferromagnetic microsphere levitated and confined by Meissner effect in a superconducting trap \cite{chris, maglev}. The experimental setup has been described in detail in Ref.~\cite{maglev}. The microsphere is made of a neodymium-based alloy with density $\varrho=7.4 \times 10^3$ kg/m$^3$ and radius $R=\left(27 \pm 1 \right)$ $\mu$m (Fig.~\ref{scheme}), fully magnetized in a $10$ T NMR magnet prior to the experiment, with an expected saturated magnetization $\mu_0 M \approx 0.7$ T. It is levitated by Meissner effect inside a cylindrical well machined in a $99.95\%$-purity Pb block with $4$ mm diameter and $4$ mm depth. The Meissner surface currents induced by the magnetic microsphere, combined with gravity, provide full confinement in all spatial direction. The motion of the microsphere is detected by a commercial dc SQUID connected through a single pick-up coil placed above the levitated particle.  The pick-up coil consists of $6$ loops of NbTi wire, wound around a cylindrical PVC holder with radius $1.5$~mm, coaxial with the trap. The setup is mounted inside a magnetically shielded copper vacuum chamber filled with a variable pressure of helium gas, which is dipped in a standard helium transport dewar at $T=4.2$ K. We monitor the helium pressure in the vacuum chamber with a Pirani-Penning gauge placed at room temperature. The actual pressure at the microsphere location is then estimated by applying a correction which takes into account the thermomolecular pressure drop \cite{maglev,sydoriak}. In the low pressure limit, this can be approximated as $P/P_0=(T/T_0)^{\frac{1}{2}}$ where $P$ and $T=4.2$ K are pressure and temperature at the microsphere location, $P_0$ is the helium pressure measured by the gauge at room temperature and $T_0 \approx 300$ K.

As discussed in Ref.~\cite{maglev}, the SQUID is able to detect 5 degrees of freedom of the rigid body. Comparison with a finite element simulation allows to reliably identify $3$ translational modes and $2$ librational modes. In this work we focus on the vertical translational mode, which we refer to as the $z-$mode. For this mode the resonance frequency can be also approximately estimated by applying the image method to a magnetic dipole above an infinite plane \cite{maglev}:
\begin{equation}
  f_0=\frac{1}{\pi}\sqrt{\frac{g}{z_0}}   \label{f0}
\end{equation} 
where $z_0$ is the equilibrium height:
\begin{equation}
z_0=\left( \frac{3 \mu_0 \mu^2}{64 \pi m g }\right)^{\frac{1}{4}}
\end{equation}
Here, $\mu=MV$ is the total magnetic dipole moment with $M$ saturation magnetization and $V$ volume, $m=\varrho V$ is the mass, and $g$ is the gravity acceleration. For our microsphere we estimate $f_0=59.0$ Hz, not far from the measured value $f_0=56.8$ Hz.
A small discrepancy is entirely expected, because the image method is exact only for a magnet above an infinite superconducting plane, and cannot account for the finite size of the trap. However, the substantial agreement between the experimental frequency, the image method and finite element simulations \cite{maglev}, strongly supports the identification of the mode at $56.8$ Hz as the translational $z$-mode.

\begin{figure}[!ht]
\includegraphics[width=8.6cm]{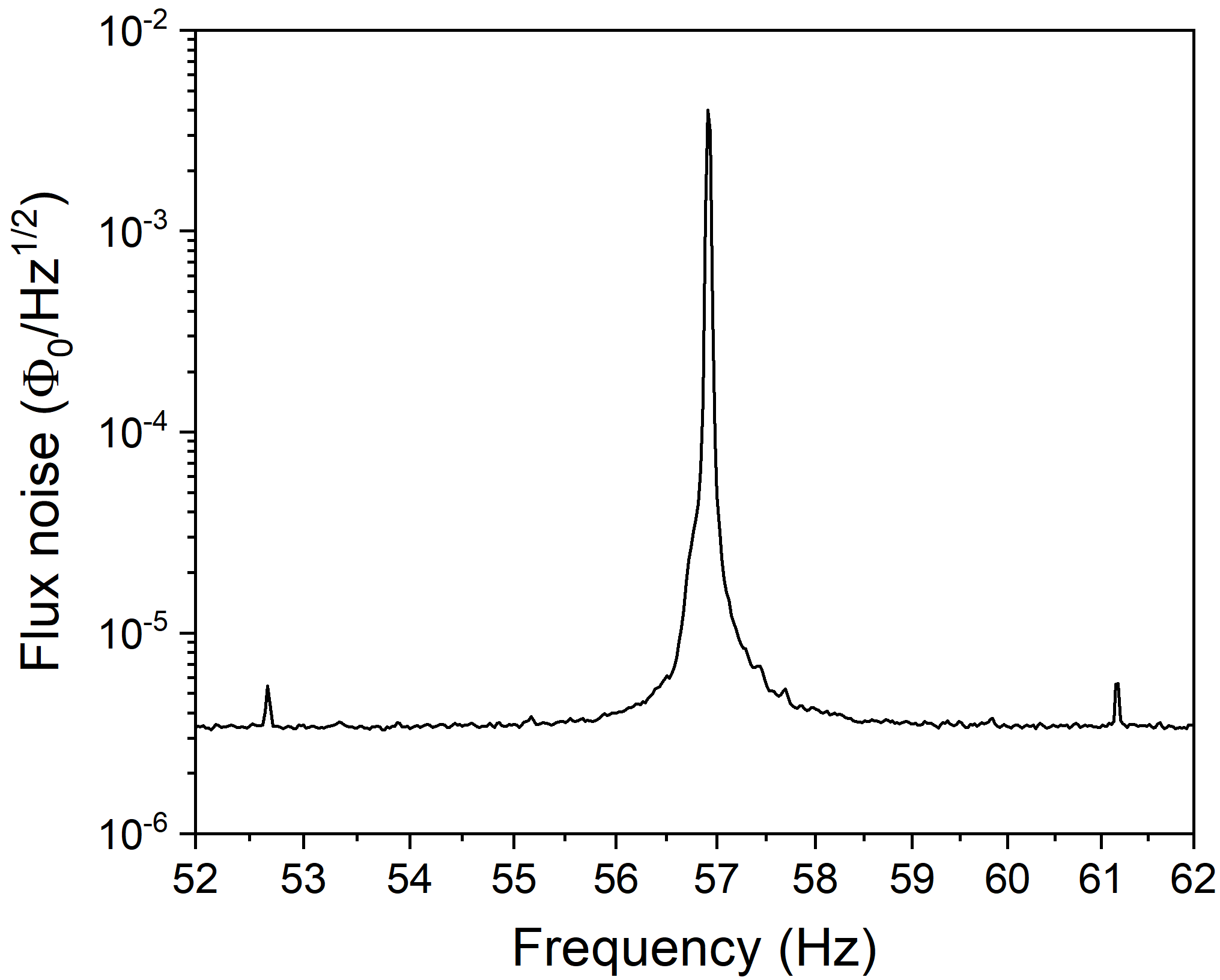}
\caption{Typical power spectrum of the $z-$mode over $12$ hours, acquired at a pressure $P=5.6 \times 10^{-5}$ mbar. The two small satelites correspond to a nonlinear mixing with an horizontal mode at $4.3$ Hz}  \label{spectrum}
\end{figure}

\section{Experimental results}
Fig.~\ref{spectrum} shows an uncalibrated spectrum of the $z-$mode, expressed in units of magnetic flux at the input of the SQUID, averaged over $12$ hours. The resonance frequency is remarkably stable over time, featuring only small amplitude-dependent shifts due to anharmonicities in the trapping potential. During the measurements relevant to this paper these shifts are always smaller than $1$ Hz.

\begin{figure}[!ht]
\includegraphics[width=8.6cm]{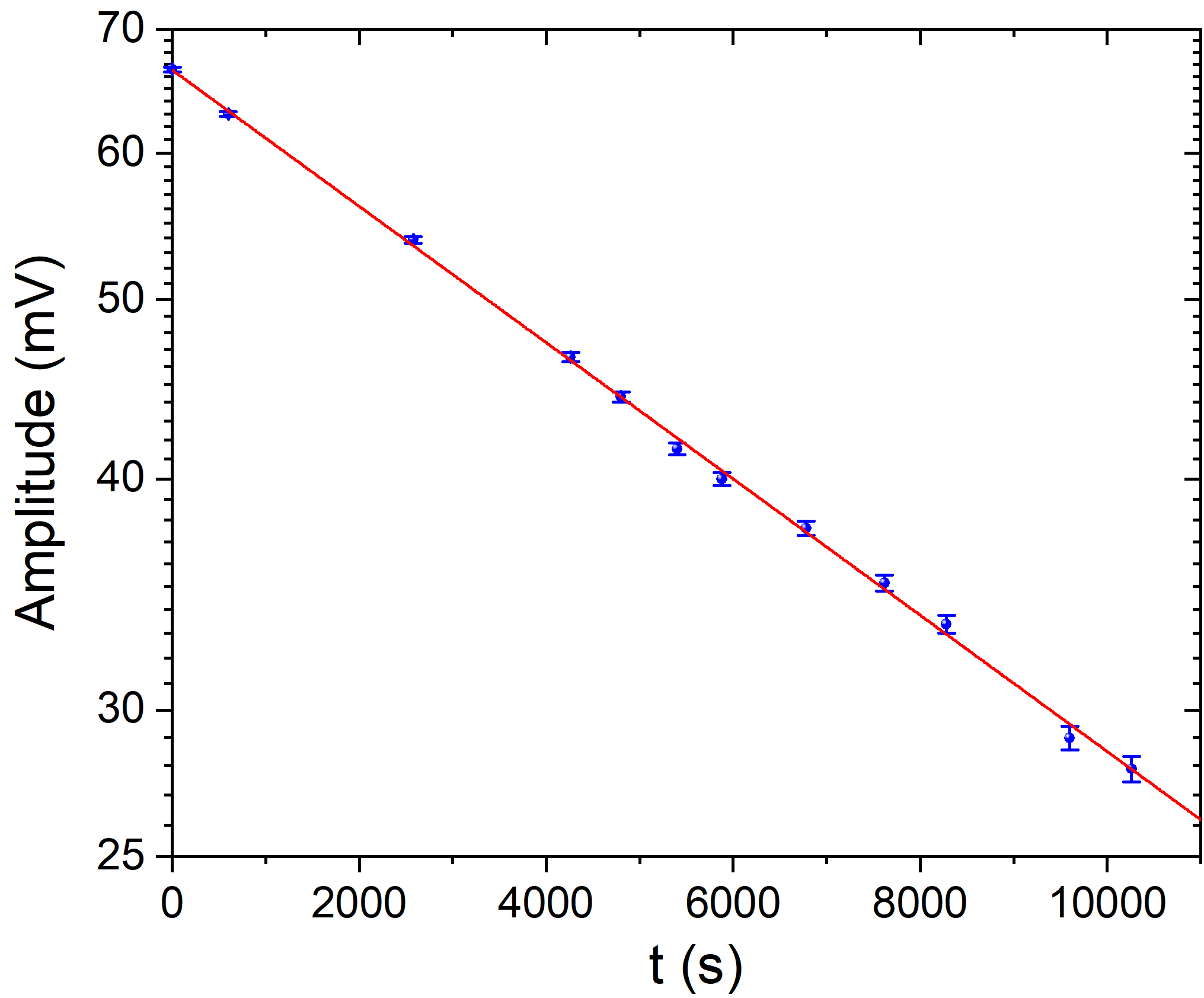}
\caption{Ringdown measurement performed at $P=1.35 \times 10^{-5}$ mbar. The weighted exponential fit provides the amplitude decay time $\tau=\left(1.19 \pm 0.01 \right) \times 10^4$ s.}  \label{decay}
\end{figure}

Fig.~\ref{decay} shows a ringdown measurement of the $z-$mode. The mode is excited by sending an ac current of the order of $1$ mA through a single loop excitation coil wound on the pick-up coil holder. After excitation, we monitor the ringdown by means of a lock-in amplifier with reference frequency $f_r$ set close to the actual amplitude dependent resonance frequency $f_0$. Before any amplitude measurement we precisely adjust $f_r$ to $f_0$ to better than $1$ mHz by nulling the phase drift rate. The error bar on each point is calculated by adding in quadrature the mean amplitude of the peak when it is dominated by noise.
The data in Fig.~\ref{decay} correspond to the lowest damping effectively measured in the experiment, $\gamma = 2/\tau=\left(1.68  \pm 0.02 \right) \times 10^{-4}$ s$^{-1}$. Note that it is common in literature to report the dissipation in terms of a linewidth in Hz \cite{pontin}, which in our case is given by $\gamma/2 \pi= \left( 26.7 \pm 0.4 \right)$ $\mu$Hz. 

\begin{figure}[!ht]
\includegraphics[width=8.6cm]{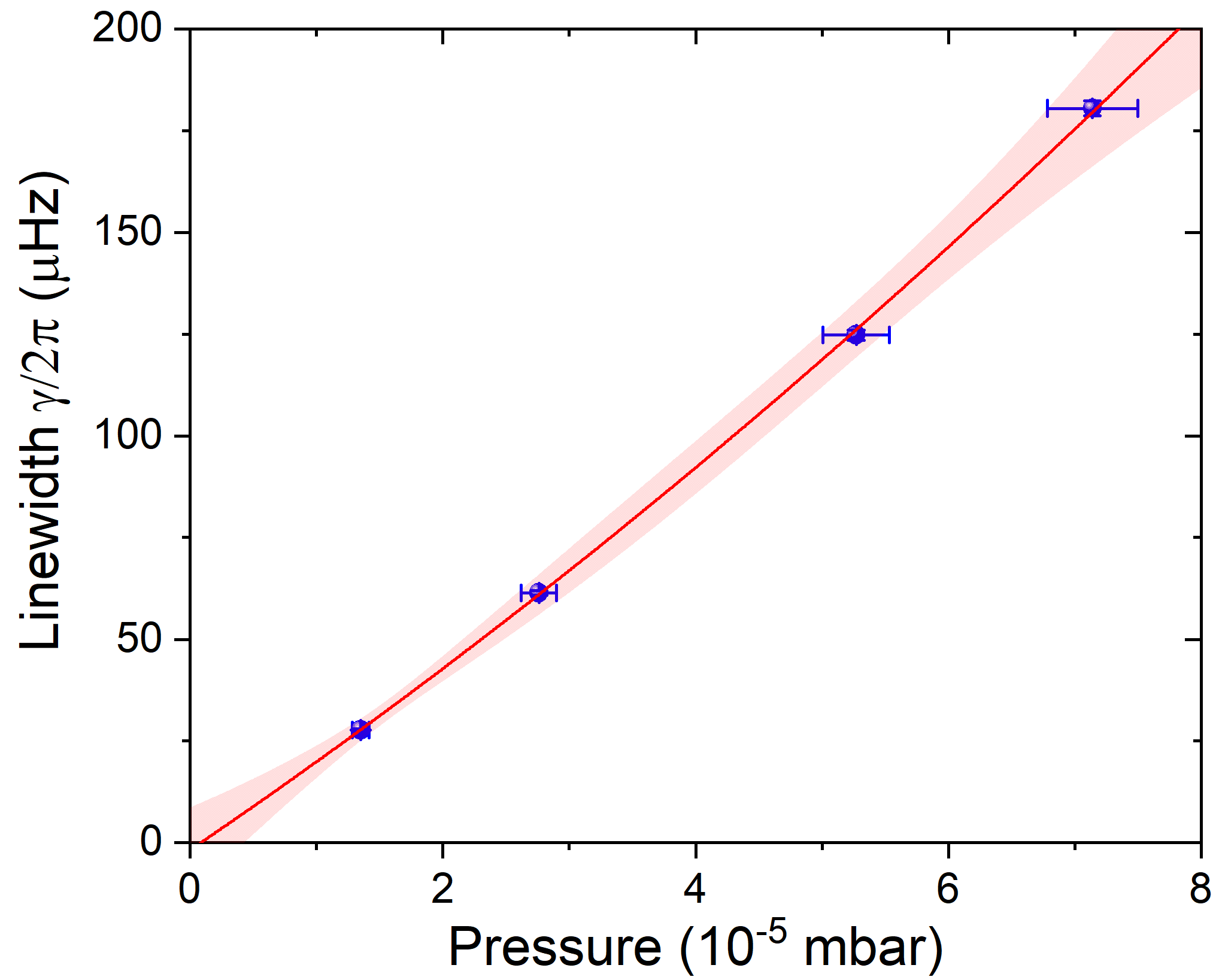}
\caption{Linewidth as a function of the pressure for the $z$-mode. A second order polynomial fit is shown, together with the 90 $\%$ confidence bands.}  \label{dampingvsP} 
\end{figure}

In Fig.~\ref{dampingvsP} we report the linewidth as a function of the pressure for the $z$-mode. The uncertainty is dominated by the error in the determination of pressure. We observe approximately a linear dependence on $P$, as predicted by standard gas damping models \cite{epstein}. We note that the correction for the thermomolecular effect is accurate only in the low pressure limit but breaks down at higher pressure \cite{maglev}. We take into account a possible deviation from linearity in the data by adding a quadratic term in the fitting function. 
The second order polynomial fit is shown in Fig.~\ref{dampingvsP} together with the $90\%$ confidence intervals. The linear term is $\left( 2.1\pm 0.1 \right)$ Hz/mbar and can be directly compared with the gas damping prediction, given by \cite{epstein}:
\begin{equation}
  \gamma/2 \pi=\frac{1}{\pi} \left(1+\frac{8}{\pi} \right)\frac{P}{\varrho R v_{\mathrm{th}}}
\end{equation}
where $v_{\mathrm{th}}=\sqrt{8 k_B T/\pi m_g}$ is the mean thermal velocity of the gas and $m_g$ is the molecular mass of helium. By inserting the numerical values we obtain $\gamma/(2 \pi P)=1.9$ Hz/mbar, in fair agreement with the experimental value.
From the confidence intervals we infer a linewidth at zero pressure $\gamma_0/2 \pi <9$ $\mu$Hz at 90$\%$ confidence level. We will use this value as an upper limit on a possible dissipation arising from collapse models. 

\begin{figure}[!ht]
\includegraphics[width=8.6cm]{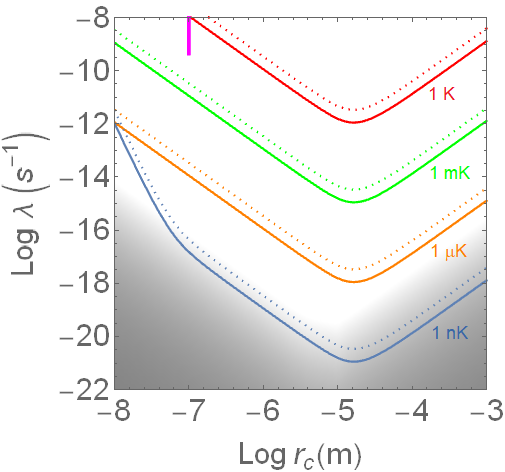}
\caption{Exclusion plot for the dCSL model in the $\lambda-r_c$ plane. The solid curves correspond, from top to bottom, to $T_c=1$ K (red), $T_c=10^{-3}$ K (green), $T_c=10^{-6}$ K (orange), $T_c=10^{-9}$ K (blue). Dotted lines are shown as reference to show the bound from the lowest actually measured dissipation rather than the extrapolated one: from top to bottom $T_c=1$ K (red), $T_c=10^{-3}$ K (green), $T_c=10^{-6}$ K (orange), $T_c=10^{-9}$ K (blue). The shaded gray region is conventionally considered unnatural for the CSL model, as it would not guarantee an effective collapse of macroscopic quantum superpositions \cite{toros}. The vertical bar represents the enhanced values for $\lambda$ at $r_c=10^{-7}$ m proposed by Adler \cite{adler}.}  \label{dCSL} 
\end{figure}

\section{Discussion} 

\subsection{The dCSL model}

Our experimental data can be used to exclude the regions of the dCSL parameter space which predict a dissipation larger than the one measured in the experiment. Fig.~\ref{dCSL} shows a family of curves in the $\lambda-r_c$ plane, each one corresponding to a fixed temperature $T_c$. The region above each curve is experimentally excluded by our experiment at $90$~$\%$ confidence level. For reference we also show as dotted lines the bounds that are obtained by considering the actually measured dissipation rather the extrapolated one at zero pressure. The gray region of parameter space is conventionally considered unnatural for the CSL model, as parameters well inside this region would not guarantee an effective collapse of macroscopic superpositions \cite{toros}. In other words the CSL model would no more accomplish its original scope. For $r_c=10^{-7}$ m the gray region is equivalent to the initial value for $\lambda$ proposed by Ghirardi {\it et al.} \cite{CSL}. The vertical bar represents the enhanced values for $\lambda$ proposed by Adler \cite{adler}.

Clearly, our approach is particularly sensitive to low values of $T_c$, as these imply large values of dissipation. For $T_c\approx 10^{-9}$ K, the blue curve in Fig.~\ref{dCSL}, almost the entire natural parameter space of CSL is excluded. We also note a new feature on the left side of the $T_c\approx 10^{-9}$ K curve, with the slope which becomes much steeper, from a $\sim r_c^{-2}$ to a $\sim r_c^{-6}$ dependence. This corresponds to the transition from weak dissipation $\chi < 1$ to strong dissipation $\chi>1$.
Our results can be compared with a similar experiment performed with a nanoparticle in a Paul trap \cite{pontin}. In particular, our bounds are more stringent for $r_c >2 \times 10^{-7}$ m.
If we compare our bounds on dCSL, based solely on dissipation, to the bounds which can be inferred from noise measurements \cite{vinanteCSL1,vinanteCSL2, vinanteCSL3, lisa}, we notice that the latters have only recently been able to exclude the enhanced values for $\lambda$ proposed by Adler \cite{vinanteCSL3}. Therefore, bounds on dCSL inferred from dissipation for $T_c \lesssim  10^{-3}$ K are already much stronger than bounds inferred from noise. 

As a guide to future experiments it is useful to discuss some scaling properties of the exclusion curves for different values of radius and dissipation. This can be done by a closer inspection of Eq.~(\ref{gamma1}). We find that the minimum of the exclusion curve is achieved for $r_c \approx 0.6 R$, and the value of the curve at minimum is proportional to $R^{-1}$. For instance, for a microsphere with the same density and mechanical dissipation but with radius $R$ smaller by a factor of 100, the minimum of the exclusion curve would be shifted to $r_c \approx 10^{-7}$ m and the value at minimum would be around $10^{-10}$ Hz for $T_c=1$ K. This would allow to falsify Adler's interval for the most plausible value of $T_c$ according to the proponents of dCSL. To achieve the same result with the micromagnet used in our experiment we would need to reduce the dissipation by a factor of $100$.

\begin{figure}[!ht]
\includegraphics[width=8.6cm]{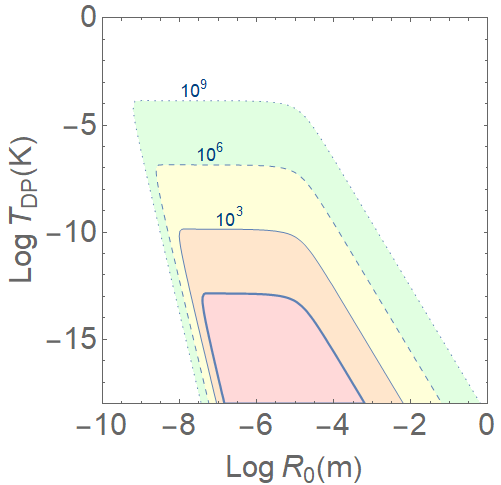}
\caption{Exclusion plot for the dDP model in the $T_\text{DP}-R_0$ plane, by assuming $m_a=4 \pi/3 \varrho R_0'^3$. The inner light red region, delimited by a thick solid line, is excluded by our experiment. The outer regions (orange, yellow, green) respectively delimited by thin solid, dashed and dotted lines, would be excluded by a reduction of the experimental dissipation rate by a factor $10^3, 10^6, 10^9$ respectively. The dDP model with $T_{\text{DP}}=1$ K would predict a dissipation rate $10^{13}$ times smaller than the observed one. The cut-off at $R_0 \approx 10^{-5}$ m is determined by the size of our particle.}  \label{dDP} 
\end{figure}

\subsection{The dDP model}

For the dDP model, using for $m_a$ in Eqs.~(\ref{chiDP}) and (\ref{gammaDP}) the mean nuclear mass, we find that our experiment does not provide any exclusion in the uniform matter limit Eq.~(\ref{etaDP1}). In the granular matter limit Eq.~(\ref{etaDP2}), it formally provides an exclusion region, but this corresponds to unphysical parameters $R_0<10^{-23}$ m. In fact, the Di\'{o}si-Penrose model is nonrelativistic, and this assumption breaks down for $R_0\ll 10^{-15}$ m. Furthermore, in \cite{smirne1} it has been pointed out that already for $R_0=10^{-15}$ m a dissipative extension of the DP model would lead to instability of nuclear matter.  

However, we find a significant exclusion by making a different choice for the reference mass $m_a$ which appears in the expressions of $\chi_\text{DP}$ and $\gamma_\text{DP}$. Specifically, we can take as elementary entity for the dDP mechanism a sphere of radius $R_0'=R_0 \left( 1+\chi_\text{DP} \right)$, i.e. $m_a=\frac{4 \pi}{3} \varrho R_0'^3$. This choice is motivated by the fact, apparent from Eq.~(\ref{etaDP1}), that the collapse mechanism is coherent within a sphere of radius $R_0'$, that is $\eta_{\text{DP}}\propto m^2$, while it scales linearly with the mass for $R>R_0'$. In other words, an object smaller than $R_0'$ behaves as a single particle of mass $m$, meaning that the physics of the DP collapse mechanism is suppressed below the $R_0'$ scale. Under this assumption, the excluded region in the $T_\text{DP}-R_0$ parameter space, that is the region where the predicted dissipation is larger than the observed dissipation, is shown in Fig.~(\ref{dDP}).
We note that the excluded region extends up to a temperature $T\approx 10^{-13}$ K. A given reduction of the measured dissipation rate $\gamma$ would shift the bound up by the same factor. Therefore, we are roughly $13$ orders of magnitude off from excluding the dDP model with $T_\text{DP}=1$ K. The upper bound on $T_\text{DP}$ does not depend on the size of the object $R$. For larger $R$ and same $\gamma$ we would however observe a shift of the high $R_0$ cutoff, which is located at $R_0 \approx R$.

\subsection{The CGF model}
\begin{figure}[!ht]
\includegraphics[width=8.6cm]{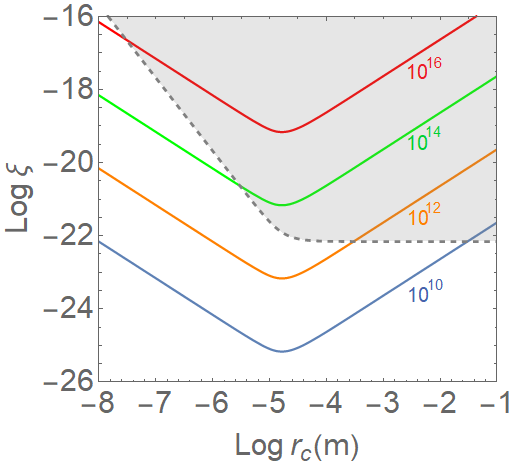}
\caption{Exclusion plot for the CGF model in the $r_c-\xi$ plane, where $r_c$ is the correlation length and $\xi$ is the magnitude of the complex fluctuations of the metric. The solid curves correspond, from top to bottom, to different correlation rate $\lambda=10^{16},10^{14},10^{12},10^{10}$~Hz. The regions above the curve is excluded by the experiment. The thick dashed gray curve correspond to the choice $\lambda =c/r_c$, and under this condition the gray region above the curve is excluded by the experiment. }  \label{CGF} 
\end{figure}

Figure~\ref{CGF} shows the exclusion plot for the CGF mode in the ($r_c,\xi$) plane, where $r_c$ is spatial correlation length and $\xi$ is the magnitude of the complex gravity fluctuation. Different curves are plotted corresponding to different correlation times. A physical insight is provided by the thick blue curve, which is obtained by Eq.~(\ref{gammaCGF2}), i.e. by assuming that temporal and spatial correlations are related by $c$. This is suggested by the fact that the gravitational field propagates at speed of light \cite{gasbarritoros}. The gray region is then excluded by our experiment. Interestingly, the order of magnitude of the probed region, down to $10^{-22}$ is comparable with the typical amplitude of the metric represented by astrophysical gravitational waves. It should be stressed that the fluctuations of the metric probed by our experiment have a nature quite different from gravitational waves: they are complex, and the correlation time is very short.

\section{Conclusion}
We have set new improved bounds on dissipative collapse models, based on measuring ultralow dissipation in a low frequency levitated micromagnet. Our data are essentially ruling out the dCSL model for collapse field temperatures of $10^{-9}$ K or lower. For the dDP model the exclusion is much weaker. By setting $m_a$ as the mass of a sphere of radius $R_0'$, we exclude field temperature $T_\text{DP}<10^{-13}$ K. We have also tested the magnitude of complex metric fluctuations suggested by the CGF model, and in particular we have set for the first time a bound on the imaginary part of the correlator of such fluctuations, directly related to dissipation. We have probed fluctuations of the metric with amplitude down to $10^{-22}$.

All data supporting this study are openly available from the
University of Southampton repository in Ref.~\cite{data}.

\begin{acknowledgments}
A.V., C.T., G.G. and H.U. acknowledge financial support from the EU H2020 FET project TEQ (Grant No. 766900), the Leverhulme Trust (RPG-2016-046), the COST Action QTSpace (CA15220) and the Foundational Questions Institute (FQXi). M.T. acknowledges financial support from EPSRC (grant N. EP/N031105/1).
\end{acknowledgments}

\appendix

\section{Calculational details for the granular limit (lattice model)}

The amplification factor for spherical particles in the regime of
tiny motional displacements, both for the CSL and DP models, has been
discussed in detail in Ref.~\cite{nimmrichter}. The dCSL and dDP models in the regime of tiny displacements have been discussed in Refs.~\cite{nobakht} and \cite{pontin}.

The amplification factor can be well understood in terms of Adler's
formula or within the homogeneous-body approximation, but the most refined
modelling is based on a lattice model (for a comparison see~\cite{toros}).
Here we discuss the extension of the the latter~\cite{nimmrichter}
to the dissipative CSL and DP models \textendash{} we will see that
most of the analysis carries over to the dissipative variants. In
particular, we focus on the regime where the effective localization
lengths, $r_{c}'=r_{c}(1+\chi)$ and $R_{0}'=R_{0}(1+\chi_\text{DP})$, are smaller than the lattice
constant $a$; we expect a linear scaling of the amplification parameter
$\eta$ or $\eta_\text{DP}$ with the mass of the system as expected from Adler's formula.
The other interesting regimes of the dissipative models for spherical
particles have been reported in~\cite{pontin}.

We consider the mass density of a spherical body:
\[
\varrho(\bm{r})=m_{a}\sum\delta(x-an_{x})\delta(y-an_{y})\delta(z-an_{z})
\]
where $a$ is the lattice number, the sum is over the values $n_{x}^{2}+n_{y}^{2}+n_{z}^{2}\leq n_{\text{max}}^{2}$,
$an_{\text{max}}$ is the radius of the body, and $m_{a}$ is the
mass of a unit cell. The Fourier transform of the mass density is
given by
\begin{equation}
\tilde{\varrho}(\bm{q})=\int d\bm{r}\varrho(\bm{r})e^{i\frac{\bm{q}\cdot\bm{r}}{\hbar}}.
\end{equation}
For later convenience we evaluate 
\begin{equation}
\vert\tilde{\varrho}(\bm{q})\vert^{2}=m_{a}^{2}\sum\sum e^{i\frac{aq_{x}\Delta n_{x}}{\hbar}}e^{i\frac{aq_{y}\Delta n_{y}}{\hbar}}e^{i\frac{aq_{z}\Delta n_{z}}{\hbar}}\label{eq:mut2}
\end{equation}
where $\Delta n_{j}=n_{j}-l_{j}$, and the double sum is over the
values $n_{x}^{2}+n_{y}^{2}+n_{z}^{2}\leq n_{\text{max}}^{2}$ and
$l_{x}^{2}+l_{y}^{2}+l_{z}^{2}\leq n_{\text{max}}^{2}$. 
\\
\subsection*{The dCSL case}

We start from
\begin{equation}
\eta=\left[\frac{\nu^{2}}{(2\pi\hbar)^{3}}\frac{1}{\hbar^{2}}\right]\int d\bm{q}\vert\tilde{\varrho}(\bm{q})\vert^{2}e^{-\frac{r_{c}^{2}(1+\chi)^{2}\bm{q}^{2}}{\hbar^{2}}}q_{x}^{2},
\end{equation}
where 
\begin{equation}
\nu^{2}=\frac{\lambda r_{c}^{3}(4\pi)^{3/2}}{m_{0}^{2}}.\label{eq:nu2}
\end{equation}
Using Eq.~(\ref{eq:mut2}) we readily find:
\begin{align}
\begin{aligned}
\eta=&\frac{\nu^{2}}{(2\pi)^{3}}\frac{\pi^{\frac{3}{2}}m_{a}^2 }{4 r_c'^{7}} \times \\
&\sum\sum\left(2 r_{c}'^{2}-a^{2}\text{\ensuremath{\Delta n_{x}^{2}}}\right)e^{-\frac{a^{2}\left(\text{\ensuremath{\Delta n_{x}^{2}}}+\ensuremath{\Delta n_{y}^{2}}+\ensuremath{\Delta n_{z}^{2}}\right)}{4 r_{c}'^{2}}}.
\end{aligned}
\end{align}
We now assume $a\gg4 r_{c}'$ such that only the terms satisfiying $\text{\ensuremath{\Delta n_{x}^{2}}}=\ensuremath{\Delta n_{y}^{2}}=\ensuremath{\Delta n_{z}^{2}}=0$
contribute; the contribution from a single sum is $m/m_{a}$. This immediately gives
\begin{equation}
\eta=\frac{\nu^{2}m_{a}m}{16\pi^{3/2}r_{c}'^{5}}.
\end{equation}
We now use Eq.~(\ref{eq:nu2}) to find
\begin{equation}
\eta=\frac{\lambda}{2(1+\chi)^{5}r_{c}^{2}}\frac{m_{a}m}{m_{0}^{2}}.
\end{equation}
\\

\subsection*{The dDP case}

We start from
\begin{equation}
\eta_{\text{DP}}=\left[\frac{G}{2\pi^{2}\hbar^{2}}\frac{1}{\hbar^{2}}\right]\int d\bm{q}\vert\tilde{\varrho}(\bm{q})\vert^{2}e^{-\frac{R_{0}^{2}(1+\chi_\text{DP})^{2}\bm{q}^{2}}{\hbar^{2}}}\frac{q_{x}^{2}}{\bm{q}^{2}}.
\end{equation}
We insert Eq.~(\ref{eq:mut2}) and, similarly as for dCSL, assume
$a\gg4(1+\chi_\text{DP})R_{0}$ such that only the terms satisfiying $\text{\ensuremath{\Delta n_{x}^{2}}}=\ensuremath{\Delta n_{y}^{2}}=\ensuremath{\Delta n_{z}^{2}}=0$
contribute. We then find 
\begin{equation}
\eta_{\text{DP}}=\frac{G}{6\hbar\sqrt{\pi}(1+\chi_\text{DP})^{3}R_{0}^{3}}m_{a}m.
\end{equation}
We obtain the same result from \cite{nimmrichter}
with the formal replacement $R_{0}\rightarrow R_0' =(1+\chi_\text{DP})R_{0}$.


\begin{thebibliography}{<99>}
\bibitem{note} This assumption is usually valid in the CSL model, because too small $r_{c}$ would lead to appreciable effects in microscopic systems. However, for sake of completeness, we report in Appendix the derivation of the parameter $\eta$ for the case $r_{c} \ll a$.
\end{thebibliography}
\end{document}